\documentclass{epl}

\usepackage{amsmath}%
\usepackage{amssymb}%
\usepackage{graphicx}
\usepackage{dcolumn}
\usepackage{bm}

\title{Super-roughening as a disorder-dominated flat phase}
\shorttitle{Super-roughening as a flat phase}

\author{S. Ares\inst{1}, A. S\'anchez\inst{1} and A. R. Bishop\inst{2}}
\institute{
\inst{1} Grupo Interdisciplinar de Sistemas Complejos (GISC, {\tt http://gisc.uc3m.es}) and
Departamento de Matem\'aticas,
Universidad Carlos III de Madrid\\ Avenida de la Universidad 30, 28911
Legan\'es, Madrid, Spain \\
\inst{2} 
Theoretical Division and Center for Nonlinear Studies, MS B258\\ Los
Alamos National Laboratory, Los Alamos, NM 87545
}%


\begin{document}

\maketitle

\begin{abstract}
We study the phenomenon of super-roughening found on surfaces growing on
disordered substrates. We consider a one-dimensional version of the
problem for which the pure, ordered model exhibits a roughening phase
transition. Extensive numerical simulations combined with analytical
approximations indicate that super-roughening is a regime of asymptotically flat
surfaces with non-trivial, rough short-scale features arising from the 
competition between surface tension and disorder. 
Based on this evidence and on previous simulations of the two-dimensional
Random sine-Gordon model [S\'anchez {\em et al.}, Phys.\ Rev.\ E
{\bf 62}, 3219 (2000)], we
argue that this scenario is general and explains equally well the
hitherto poorly understood two-dimensional case.
\end{abstract}

Understanding phase transitions in systems with quenched disorder
is a challenging issue, not only in 
physics but also in fields ranging from combinatorics to
financial market modeling.
This is the reason why much effort has been devoted in the last 
few years to this subject. However, the progress achieved so far is modest, 
and clear-cut, definitive results are rare \cite{Plischke}.
In view of this, steps towards providing such results are needed and
will be very helpful for many ongoing investigations.
One important instance where the role
of disorder remains unclear is that of roughening transitions
of crystal surfaces \cite{Weeks}: At high
temperatures, crystal surfaces are rough (the surface width
$w$ increases with the system size $L$), whereas at low temperatures
they become flat ($w$ is a constant depending only on temperature).
When crystal planes are grown on a disordered
substrate, renormalization group calculations
for the two-dimensional (2D) random sine-Gordon model
(RsGM, see below) predicted that 
the conventional roughening transition was 
replaced by {\em super-roughening} \cite{Toner}, a new phase 
transition 
from the high temperature rough phase
($w\sim\sqrt{\ln L}$) to a low temperature {\em rougher}
phase ($w\sim\ln L$). 
Subsequent work
produced a
``myriad of predictions'' \cite{Batrouni} in a few years, which
gave rise to a controversy about the nature of the super-rough
phase (see \cite{Shapir} for a review).  Based mainly on
large-scale numerical simulations or exact optimization results
\cite{Riegercom,Marinari,Lancaster,Zeng,Rieger1,Coluzzi,us1,jjfises,Rieger2},
some degree of consensus was reached that super-roughening indeed took place and
that $w\sim\ln L$ was observed. However,
recent numerical evidence from simulations by our group on the 2D RsGM 
\cite{us} has cast doubts on
the generality of that behavior as the parameters of the model
change, suggesting that the low temperature phase of the model could 
even be flat. 

With the expectation that the problem would be more amenable both
analytically and numerically, in this Letter
we study the same problem in a one-dimensional (1D)
setting. To mimic as closely as possible the 2D situation, we
need to find a 1D model whose
non-random version exhibits a true roughening transition.
The 1D sine-Gordon model does not fulfill this condition (see
\cite{us2} for a rigorous proof; see also \cite{us3} for additional
numerical results)
and therefore we must modify it. To this
end, we 
introduce and study for the first time a new model, 
rooted in
the work of Burkhardt \cite{burkhardt}, who proved that the
Hamiltonian given by (periodic boundary conditions on a lattice with 
$N$ sites are used)
$
{\cal H}=\sum_{i=1}^{N}\Big\{J|h_{i+1}-h_i|+U(h_i)\Big\},
$
with $U(x)$ being a square well potential, $U(x)=-U_0$ for $0<x<R$,
$U(x)=0$ otherwise, and with {\em the heights $h_i$ restricted to
be non-negative} (impenetrable substrate) has a roughening transition
(see \cite{CW} for analogous discrete models). We stress that this is
a true thermodynamic phase transition, which 
is perfectly possible in 1D (see \cite{us4} for a thorough
discussion of 1D phase transitions).
Building on these results, we propose the following
Hamiltonian as our basic model (hereafter called Burkhardt-RsGM or
B-RsGM):
\begin{equation}
\label{ourh}
{\cal H}=\sum_{i=1}^{N}\Big\{\frac{1}{2}(h_{i+1}-h_i)^2+
[V(h_i-h^{(0)}_i)]\Big\},
\end{equation}
where $V(x)\equiv V_0(1-\cos x) +U(x)$,
$h^{(0)}_i$ are uncorrelated random variables uniformly distributed
in the range $[0,h_{max}]$, and the impenetrable substrate restriction
now reading $h^{(0)}_i\leq h_i\leq \infty$.
We note that the original proposal of Burkhardt had a surface tension term
given by $|h_{i+1}-h_i|$. Instead, model
(\ref{ourh}) follows the spirit of the
RsGM, including surface tension 
as the square of a discrete 
gradient term \cite{nota1}
and a periodic potential
favoring multiples of $2\pi$ for the heights $h_i$, mimicking the growth
of crystal layers. 
The 1D RsGM is recovered 
by setting $U_0=0$ and allowing for negative heights. 

We have not been able to solve the statistical mechanics 
of the {\em ordered} B-RsGM (i.e., $h_{max}=0$) exactly.
Therefore, we have resorted to
numerical simulations to check whether the roughening transition in
the original Burkhardt model carries over to our modified version.
To this end, we have used parallel tempering Monte Carlo \cite{newman,iba}.
Representative configurations at a given temperature are generated with
a heat bath algorithm \cite{toral}, 
full details of which can be found in \cite{us3}.
The parallel tempering algorithm
then considers simultaneous copies of the system at different
temperatures, allowing exchange of configurations between them.
This is particularly efficient for low
temperature configurations, which are most susceptible to being trapped
in metastable regions, particularly in the disordered case.

\begin{figure}
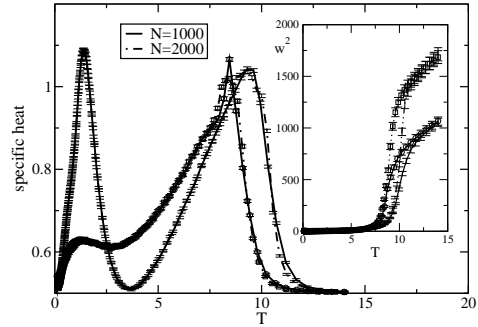

\onefigure[width=6.2cm]{tdf.eps}
\caption{Specific heat vs temperature for the B-RsGM for
two system sizes, as indicated.
Inset: Roughness vs temperature.
Lines with and without 
symbols correspond to the disordered and ordered case, 
respectively.  Error bars correspond to the
thermal average.}
\label{fig:tdf}
\end{figure}

\begin{figure}
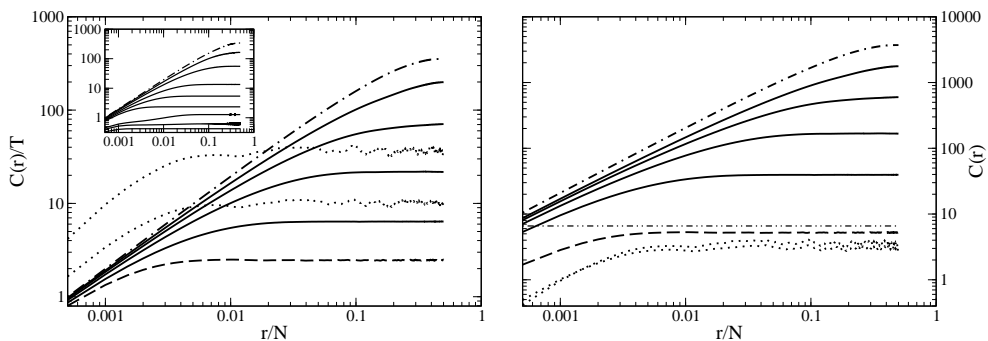

\onefigure[width=13.0cm]{letter3.eps}
\caption{ Height difference correlation function
scaled by the temperature $T$ (left panel) and without scaling
(right panel). Temperatures are $T=10.50$ (dot-dashed), $T=8.84, 8.41,
7.70, 6.17$ (solid), $T=2.13$ (dashed), and $T=0.301, 0.0986$ (dotted).
Inset: scaled height difference correlation function without
disorder.
Temperatures are $T=14.0$ (dot-dashed) and
$T=10.0, 9.34, 8.18, 6.90, 3.99, 0.995, 0.774,$ and
$0.0981$ (solid).
The double-dot-dashed horizontal line indicates
$C(r)$ for a surface exactly locked to the disorder, $C^{(0)}(r)$
 (see text).}
\label{fig:cor}
\end{figure}

The results of our simulations of the B-RsGM without disorder are
summarized in Fig.\ \ref{fig:tdf} for the specific
choice $V_0=1$, $U_0=2$, $R=2\pi$ (to have the same length as the
periodicity of the cosine term), and $h_{max}=2\pi$
(as in the standard 2D RsGM); other values give qualitatively the
same results. Our systems have sizes
in the range between 500 and 2000 sites. As can be seen from the
plot, the specific heat presents a clear jump at a temperature
$T_R^{ord}$ around 10.3 in our units (the peak at lower temperatures
corresponds to a well-known Schottky anomaly, see, e.g., \cite{us} 
and references therein). At the same temperature, the
width or roughness, defined as $w^2\equiv
\langle [h_i-\langle h_i\rangle ]^2\rangle$ with $\langle\cdots\rangle$
denoting thermal averages,
jumps from values close to zero to 
large, system-size dependent
values, indicating the onset of the rough phase. This
is further confirmed by the height difference correlation function,
$C(r)\equiv\langle\sum_i [h_i-h_{i+r}]^2\rangle/N$,
as shown in
the inset of the left panel of Fig.\ \ref{fig:cor}. Note that
the correlation function is scaled by the temperature,
aiming to identify the onset of
the rough phase as the point at which the curves for different
temperatures collapse and become
simply proportional to $r$ (the position along the chain).
{}From both plots we locate the transition at $T_R^{ord}=10.3$; actually,
there is a gap in the plot of $C(r)/T$ at $T_R^{ord}$, and for temperatures
slightly above this value the curves go discontinuously to the
collapsed, high temperature one.
In this rough phase, the model behaves effectively
as though
$U_0=V_0=0$ (in other words, as the Gaussian model).
All these features are exactly
what we expected from the theoretical results of the original
Burkhardt model \cite{burkhardt}, and make us confident that the
B-RsGM presents a roughening phase transition in the absence of
disorder, very much like the 2D RsGM. 
Further confirmation of the existence of this phase transition and of $T_R^{ord}$
can be obtained by analytical calculations based on a continuum approach 
leading to a pseudo-Schr\"odinger equation (following \cite{schneider}), as 
well as from numerical evaluation of an exact transfer matrix 
calculation \cite{else2}.

\begin{figure}
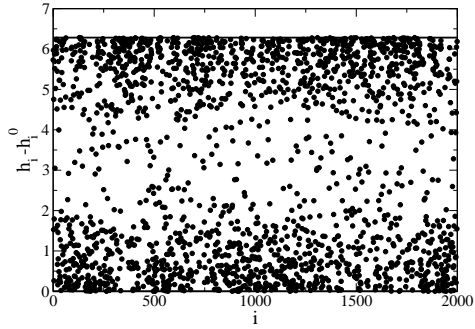

\onefigure[width=6.2cm]{height.eps}
\caption{ Heights above the substrate, $h_i-h_i^{(0)}$,
vs position along the chain. Temperature is
$T=0.0986$, system size $N=2000$.}
\label{fig:hei}
\end{figure}

\begin{figure}
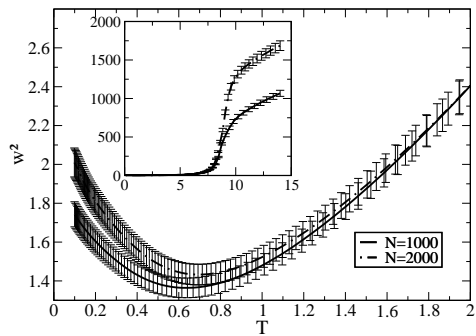

\onefigure[width=6.2cm]{rugos.eps}
\caption{Roughness as a function of temperature for the disordered model.
Inset: same for a larger range of temperatures.
}
\label{fig:rou}
\end{figure}

Having established the existence of a roughening transition
in the ordered B-RsGM, we can now proceed to
discuss how the disorder changes this transition.
Below, we present results 
of single realization simulations; we have checked
that these are typical by running several simulations with
different quenched disorder for each case. Fig.\ \ref{fig:tdf} and 
the left panel of
Fig.\ \ref{fig:cor} show that the roughening transition of the
ordered model is basically unchanged (although the Schottky peak 
is much less pronounced, due to the distortion of kinks by the
disorder); the value of $T_R$ does
change, however, and it is now located at $T_R=9.3$.
Furthermore, as the temperature is lowered and reaches $T_D=2.1$,
$C(r)/T$ begins to increase while maintaining a
finite correlation length (that implies a flat phase with roughness
independent of the system size for sizes larger than the
correlation length).
This increase arises from the fact
that, as seen in the lower panel with the temperature scaling 
removed, $C(r)$ approaches a constant function,
independent of $T$, as $T$ goes to zero, and hence scaling by
$T$ leads to an increase of the function.
The 1D character of our model allows us to study large 
systems and ensure that $C(r)$ indeed has finite range. The value
of $T$ at which this change of behavior takes place can be related
to the heights taking values close to the disorder ones: In the absence
of surface tension (the height differences term),
the minimum of ${\cal H}$ would be $h_i=h^{(0)}_i$
for all sites $i$; the correlation function for such a
configuration can be immediately found analytically, leading to a
constant value $C^{(0)}(r)=2w^2$ {\em independently of the spatial
dimension}, with $w^2\equiv\langle
(h_i^{(0)})^2\rangle-\langle h_i^{(0)}\rangle^2= h_{max}^2/12\equiv C_D$ being the squared
roughness. As can be seen from Fig.\ \ref{fig:cor}, the scaled
$C(r)$ begins to increase precisely when its asymptotic value
coincides with $C^{(0)}(r)$. This strongly suggests that at 
$T_D$ and below, the surface approaches 
the values of the disorder, while smoothed at
short scales by surface tension.
Figure \ref{fig:hei} confirms this expectation by showing
that for a majority of sites $h_i$ is close to $h^{(0)}_i$ or 
to $h^{(0)}_i+2\pi$ ($h_i$ is contained within that range at low
temperatures because of Burkhardt's square well). 
It is clear that the effect of
surface tension smooths the surface and leads to 
a global value for the roughness which is lower
than the corresponding to the disorder 
values.
In fact, by analyzing the
overdamped equations of motion for ${\cal H}$ linearized around
the substrate, it is straightforward to show that $C(r)$ goes
asymptotically to a nonzero constant $C_0$ at low temperatures, in
agreement with these numerical simulations (further details will
be given elsewhere \cite{else}). Interestingly, at temperatures as low as that
in Fig.\ \ref{fig:hei}, the roughness is slightly larger than at
higher temperatures, as seen in Fig.\ \ref{fig:rou}. This surprising
result arises, in our view, from this competition between
surface tension and disorder, which for almost zero temperatures
seems to favor the latter.

From the evidence summarized above, we conclude
that our 1D model shows no 
signatures
of super-roughening. Instead,
the picture that emerges from this work is that of a usual
roughening transition, with temperature $T_R$ lower than the
ordered case, followed at temperatures of the order or below
$T_D<T_R$ by a crossover to a disorder dominated but otherwise
flat phase. In this last regime, the surface is located close to
the 
values imposed by the disorder (modulo $2\pi$). This behavior
leads to a height-difference correlation function,
$C(r)$, that is asymptotically flat, but increasing at short
distances due to surface tension, up to a correlation length of
the order of 10-20 lattice units. At larger scales, $C(r)$ is well
described analytically by assuming that the surface height takes values close
to the disorder. Furthermore, as a consequence of the competition between
disorder and surface tension, the width exhibits a non trivial
behavior with a minimum at a low but nonzero temperature.

We believe that our conclusions are relevant in a much 
broader context than the 1D model we have described. As our model
has a true roughening phase transition, it is reasonable to 
consider it in the related context of the 2D
problem.
In the following, we will discuss the results on our 1D model we have
presented above together with our previous, extensive numerical 
work on the 2D RsGM reported in \cite{us}. 
We encourage the reader to consult \cite{us} as the 
plots, results and conclusions in that paper, which would be too lengthy to 
repeat here, clearly suggest that our interpretation is consistent 
with the available facts about the 2D RsGM. Indeed, we
believe that the abundant but often 
contradictory evidence available regarding the 2D RsGM 
can be understood within the scheme proposed here. 
The key point is that those 2D RsGM simulations
\cite{us} are perfectly consistent with our scenario:
They show
 asymptotically
flat correlation functions (with constant value approximately 
given by $C_0$) and non-monotonic roughness,
especially for strong potentials (large values of $V_0$; 
cf.\ Fig.\ 12 in \cite{us}).
It 
is true, however, that for
values of $V_0$ of the order of unity a squared logarithm behavior
is found for $C(r)$. 
In the interpretation stemming from the results reported above,
this could be due to two factors: First, for such values of $V_0$,
$T_R \gtrsim T_D$, and hence the flat phase of the ordered model
is not observed; instead, in the corresponding 
$C(r)/T$ plots in \cite{us} all 
that appears is that $C(r)/T$ increases with decreasing $T$ because,
as discussed above, $C(r)$ has become independent of $T$ and 
governed by the disorder, with the appearance of super-roughening.
Second, in 2D the surface tension influence is much larger than in
1D because of the increased number of neighbors; as a consequence,
the crossover in $C(r)$ from $C(r=0)=0$ to $C(r\to\infty)=C_D$ is much
more pronounced and occurs on a longer spatial scale. This is in
agreement with the observation in \cite{us1} that at $T=0$ 
$C(r)$ also exhibits a finite correlation length. 
Our interpretation is also consistent with the
result that changes in $V_0$ (also reported in
\cite{Batrouni,Riegercom,jjfises}) or in the interval of
the disorder $h_{max}$ lead to the disappearance of the
super-rough phase \cite{us}, replaced by a phenomenology very
close to the one we are proposing.
Importantly, we have verified with the results in \cite{us} that 
the criterion based on comparison with the tensionless
$C^{(0)}(r)$ discussed above, works equally well to determine the 
entrance to the disorder-dominated regime in 2D. 
We have checked that this is the case by applying the criterion
to our results in \cite{us} (cf.\ specially Figs.\ 3(b) and 7,
in complete agreement with the 1D results).
Furthermore, the simulations in \cite{us} yielded non-monotonic 
behaviors for the total roughness, as depicted in Fig.\ \ref{fig:rou}
for the 1D model. 
Finally, another piece of evidence in favor of our proposal is 
that sudden quenches of surfaces in the 2D RsGM lead to long lived
states with $C(r)\sim\ln^3r$ and other anomalous behaviors \cite{unpub}.
This may be related to the fact that in a quench only very short scale
rearrangements of the surface are possible, leading to a shorter 
correlation length and a shorter interval for $C(r)$ to rise from 
0 to $C_D$.

Based on the above considerations, we believe that our 
central proposal,
namely that super-roughening is an effective short-scale phenomenon
arising from the
existence of a roughening transition and a crossover to a
disorder-dominated regime is very appealing. It would naturally
explain the failures of the different theoretical
approaches, the predictions of lack of universality,
and the discrepancies among the simulations \cite{Shapir}.
Thus, obtaining different numerical results is generally the case
when studying crossover phenomena, as these are very dependent on 
the details of the model and even of the simulations. On the 
other hand, the $\ln ^2 r$ behavior for $C(r)$ can indeed be found 
if the scales analyzed are restricted to a small range, such as
those reported in 
\cite{Riegercom,Marinari,Lancaster,Zeng,Rieger1,Coluzzi},
obtained on systems of at most 200$\times$200
sites. Our own simulations on systems of sizes up to $256\times 256$ \cite{us}
(1024$\times$ 1024 at zero temperature \cite{us1}) 
confirm partly that result but show indications of a crossover to 
a flat behavior at larger time scales. 
Our proposal is an alternative explanation to the currently
available results on the phenomenon of super-roughening, but cannot be
considered as the definitive answer to this problem without further
very large-scale simulations on the 2D problem. One reason for this caveat is that
the nature of the phase transition in 
our 1D model is not exactly that of the 2D model, and the models
themselves are not identical.
We have considered this difficulty, and concluded
that the many analogies between
the numerical results reported here and those
in \cite{us}
lead us to
suggest on solid grounds that the 
mechanism underlying this common phenomenology is basically the
same. 
Caution must also be taken as it
could be argued that our results carry over only to the 2D B-RsGM
and not to the 2D RsGM. 
In this respect,
as the 2D RsGM has a roughening transition of its own, the additional
Burkhardt well would only increase the transition temperature, but not
change the general picture, and therefore both models should behave similarly.
Clearly, the scenario proposed here must
be subject to further scrutiny before accepting it. 
In particular,
to confirm that our interpretation carries over to the 2D case,
it would be necessary
to carry out large scale simulations (systems of size at least 
1000$\times$1000
at finite temperature)
to verify whether $C(r)$
crosses over, as $r$ increases, to an asymptotic, constant value, 
signaling a flat phase.
Such simulations are
presently beyond our computing capabilities and therefore we
have not been able to verify this prediction.
We hope that this
work encourages numerical work along this line to definitively
settle this long-standing issue. 
Progress in this direction would 
influence a much broader field, namely 
phase transitions in disordered systems, hence the relevance and interest of
further research on this problem.
In this context, we conclude by stressing the importance of the results
presented here. Even if they were only relevant in 1D and the scenario were 
not (as we believe it is) applicable to 2D super-roughening, our results
constitute an example of a phase transition
in a 1D disordered system which can be amenable to additional analytical work and
lead to new insights on the exciting issue of disordered systems.

\acknowledgments
This work has been
supported by the Ministerio de Ciencia y Tecnolog\'\i a of Spain
through grants BFM2000-0006 and BFM2003-07749-C05-01 
(SA and AS). Work at Los Alamos is
performed under the auspices of the US Department of Energy.


\begin{thebibliography}{88}
\bibitem{nota1} In addition, the quadratic surface tension term is much easier
   and faster to simulate than the absolute value term.
\end{thebibliography}
\end{document}